# Mössbauer and Magnetic Studies of Surfactant Mediated Ca-Mg Doped Ferrihydrite Nanoparticles

Samar Layek[1,*], M. Mohapatra[2], S. Anand[2] and H. C. Verma[1]

[1]Department of Physics, Indian Institute of Technology, Kanpur-208016, India.

[2]Institute of Minerals and Materials Technology, Bhubaneswar 751016, Orissa, India

**Abstract:**

Ultrafine (2-5 nm) particles of amorphous Ca-Mg co-doped ferrihydrite have been synthesized by surfactant mediated co-precipitation method. The evolution of the amorphous ferrihydrite by Ca-Mg co-doping is quite different from our earlier investigations on individual Ca and Mg doping studies. Amorphous phase of ferrihydrite for the present study has been confirmed by x-ray diffraction (XRD) and Mössbauer spectroscopy at room temperature and low temperatures (40K and 20K). Hematite nanoparticles with crystallite size about 8, 38 and 70 nm were obtained after annealing the as-prepared samples at 400, 600 and 800$^o$C respectively in air atmosphere. Superparamagnetism has been found in 8 nm sized hematite nanoparticles which has been confirmed from the magnetic hysteresis loop with zero remanent magnetization and coercive field and also from the superparamagnetic doublet of its room temperature Mössbauer spectrum. The magnetic properties of the 38 and 70 nm sized particles have been studied by room temperature magnetic hysteresis loop measurements and Mössbauer spectroscopy. The coercive field in these hematite nanoparticles increases with increasing particle size. Small amount of spinel $MgFe_2O_4$ phase has been detected in the 800$^o$C annealed sample.

*Author to whom correspondence should be addressed

# 1. INTRODUCTION

Iron oxides doped with various ions can completely alter and/or improve their performance for specific applications such as sensors [1, 2], photo-electrochemical activity [3, 4], catalytic activity [5] and magnetic properties [6]. In our earlier studies influence of Ca(II) on cetyltrimethyl ammonium bromide (CTAB) directed nucleation and growth of iron oxides/hydroxides was presented [7]. Detailed work on effects of variation of Ca(II) content on phase formation of iron compounds, Mössbauer parameters and magnetic behavior were reported. Depending on the amount of Ca(II) doped in iron oxides, predominant phases of goethite, mixed spherical hematite and acicular goethite or pseudo cubic hematite nanoparticles were obtained. Mössbauer studies had detected small amounts of impurity phases which were not detected in their respective XRD patterns. The VSM studies had revealed that the coercivity values were dependent on the hematite content of the sample. The sample synthesized by using 0.05M Ca(II) and 1M Fe(III) having mono dispersed primary pseudo cubic nano hematite particles (150-200 nm) gave high value of coercivity (≈1255 Oe). It was established that nonmagnetic calcium doping could influence the magnetic properties. In another investigations Mg(II) doped sample was synthesized following the same procedure using 0.05M Mg(II) and 1M Fe(III) solution. The as-prepared sample which primarily contained spherical shaped 80 nm crystalline hematite particles in combination with small amounts of amorphous ferrihydrite showed noticeable hysteresis loop with large coercive field ≈ 1562 Oe [8]. With a view to investigate the synergistic effect of doping of both Ca(II) and Mg(II) on iron phase formation and magnetic behavior of synthesized sample, the present studies were taken up. The results obtained were completely different from those expected. Formation of a combination of spherical and pseudo-cubic hematite particles was expected with doping of both Ca(II) and

Mg(II) ions, but it was surprising to observe the formation of a totally amorphous phase. The XRD, Mössbauer and magnetic studies of as synthesized and sample calcined at 400, 600 and 800°C are discussed in the present investigation.

## 2. EXPERIMENTAL DETAILS

### 2.1. Material Synthesis

The experimental procedure followed was same as given in earlier communications for single ion doping [7, 8]. The chemicals used for the synthesis of Ca-Mg doped iron oxide sample were: Fe(NO$_3$).9H$_2$O, CaCl$_2$ (fused), cetyltrimethyl ammonium bromide (CTAB) and sodium hydroxide of E-Merck, India and MgCl$_2$.6H$_2$O of Ranbaxy, India. 100 mL of mixed solution containing 1M Fe(III), 0.05M Ca(II) and 0.05M Mg(II) was taken in a beaker. To this 5 mL of 10 % CTAB solution was added and the solution was stirred for 2 h followed by pH adjustment to 10.0 with slow addition of 1 M NaOH. The contents were stirred for a period of 24 h and then filtered through a G4 frit crucible. Chloride and nitrate ions were removed by thoroughly washing the precipitate with distilled water. The Fe, Ca and Mg contents in the filtrate were analyzed and it was observed that iron was quantitatively precipitated whereas 91% for Ca(II) and 98% for Mg(II) precipitated under the employed conditions. The precipitate was then transferred to a beaker and the volume was made up to 500 mL with distilled water. The contents were aged at 60°C for 12 h in an air oven. After aging, the solid was separated by filtering through a G4 crucible, washed thoroughly with distilled water and dried in an oven at 100°C for 24 h. The as prepared sample was annealed at 400, 600 and 800°C for 4 hours each inside a programmable box furnace at a heating rate of 100 °C per hour at air atmosphere.

**2.2. Chemical Analysis and Characterization Tools**

Wet chemical method was applied for estimation of iron by taking a weighed amount of sample and subjecting to hydrochloric acid digestion. Iron was analyzed volumetrically [9] following standard procedure and other cations were analyzed by Atomic Absorption Spectrophotometer (AAS, Model Perkin Elmer Analyst 200, USA). As-prepared and all annealed samples were characterized by X-ray diffraction on a Seifert Iso-Debyeflex 2002 Diffractometer using Cr-Kα radiation in order to check the phase purity and calculate average crystallite size. Transmission electron microscope (TEM, FEI, TECNAI $G^2$ 20, TWIN) operating at 200 kV, equipped with a GATAN CCD camera was used to study the surface morphology of the sample. Mössbauer spectra of the as-prepared and calcined samples were recorded at room temperature using a standard constant acceleration Mössbauer spectrometer in the transmission geometry. A 25 mCi $^{57}$Co embedded in Rh-matrix was used as the radioactive Mössbauer source during the experiment. The collected experimental data has been fitted with a least squares computer based program by considering Lorentzian line shape of the spectrum. All the parameters have been calculated by taking the Mössbauer spectrum of bcc iron as the calibration. Low temperature (40 K and 20 K) Mössbauer spectrum was collected in a low temperature cryostat using $^4$He closed cycle refrigerator. Room temperature magnetization was measured for the as prepared and calcined samples as a function of applied magnetic field (hysteresis loops) using a vibrating sample magnetometer (VSM, ADE Technologies, USA) upto the highest applied field of 1.75 T.

**3. RESULTS AND DISCUSSIOS**

**3.1. Chemical analysis and characterization of as prepared Ca-Mg doped sample**

The chemical analysis of the sample showed that it contained 55.83% Fe, 1.81 % Ca and 1.17 % Mg. The low iron content of the sample clearly indicates that the sample is not hematite which should contain 69.9% Fe. Considering the %ages of Ca and Mg, the sample should have contained ~66% Fe content in case of formation of hematite. The XRD pattern of as synthesized sample is given in Figure 1. There are no crystalline phases and the sample is totally amorphous. Results from earlier work on no doping and individual Ca or Mg doping together with present data are given in Table 1. This is to compare the formation of different phases of iron in presence /absence of CTAB and/or Ca(II), Mg(II) ions to highlight the role of CTAB, Ca(II) ions, Mg(II) ions and a combination of Ca(II) and Mg(II) ions.

From Table 1, it is observed that all the reagents i.e., CTAB, Ca(II) and/or Mg(II) have definite role in the formation of iron phases. In spite of adding same amounts of Ca(II) or Mg(II) there is difference in their %age in the as prepared samples due to difference in the phases and Fe(III) content of the product formed. In the absence of any surfactant or additional ion, crystalline 6-line ferrihydrite was formed. The mediation of surfactant directed the conversion of ferrihydrite to goethite part of which underwent rearrangement/de-hydroxylation to give a mixture of ferrihydrite, goethite and hematite. The addition of Ca(II) or Mg(II) for the mentioned concentration in Table 1 gave primarily nano hematite as the major phase with difference in particle morphology and sizes. This had indicated that these ions either accelerate the conversion of goethite to hematite or directly convert ferrihydrite to hematite. But when same concentrations of Ca(II) and Mg(II) were added together, the product obtained was devoid of hematite. Figure 1 shows XRD pattern of this sample and one can see that the product is completely amorphous in nature. It is really very difficult to explain. It has been suggested that $Ca^{2+}$ gets adsorbed on iron surface and does not get into the lattice. This is supported by the XRD results where d-values are found to remain same on $Ca^{2+}$ doping. $Ca^{2+}$ content directs the iron oxide/oxy-hydroxide phase formation and its transformation. The non inclusion of the $Ca^{2+}$ in the lattice may be due to the fact that $Ca^{2+}$ and $Fe^{3+}$ have large difference in ionic radii and also their preference for coordination modes are different (high spin $Fe^{3+}$ prefers 6-coordination, while $Ca^{2+}$ is found in a range of coordination environments, commonly 8-coordination as in perovskite, $CaTiO_3$).

However, in case of Mg(II) doping the limited inclusion of $Mg^{2+}$ together with $Fe^{3+}$ in a co-precipitated phase can be expected considering the similarities in their ionic radii ($Mg^{2+}$, crystal radius, 6-coordination 0.860 Å, as compared to $Fe^{3+}$, 0.785 Å). $Mg^{2+}$ has been observed to be accommodated deeply into the structure of other oxides [12]. Difference of 0.02A° in the *d*-spacing values due to Mg-incorporation in the hematite matrix was observed [8]. A preferential orientation of the crystal plane was observed as the relative intensities of (104) and (110) peaks of hematite were observed to be less than the reported JCPDS (card no. 79-0007) for Mg-doped hematite.

It seems that simultaneous adsorption of Ca(II) on the surface and incorporation of Mg(II) into the iron matrix for the present sample totally restrict the conversion of amorphous ferric hydroxide phase to any crystalline form.

The TEM image of the Ca-Mg doped sample (in figure 1) shows very fine agglomerates of spherical particles in the size range of 2 to 5 nm. It may be mentioned here that by only Ca(II) doping in the mentioned concentration gave pseudo cubic monodispersed hematite particles [7] while only Mg(II) doping resulted in spherical mono dispersed hematite particles with no agglomeration.

## 3.2. X-ray diffraction of annealed samples

The room temperature X-ray diffraction patterns of annealed samples of Ca-Mg-doped hematite are presented in Figure 2. The crystallinity starts evolving in the annealed samples. For the samples annealed at 400 °C, the diffraction peaks corresponding to (104), (110), and (113) planes of hematite are identified. The full width at half maximum (FWHM) of the peaks are high due to smaller crystallite size. For the samples annealed at 600°C, all the peaks, indexed as (012), (104), (110), (113), (024), (116), (122), (214) and (300), can be assigned to the rhombohedral structure with space group *R3c* which is same as that of a pure bulk hematite sample. No other peaks from any impurity phases (like metallic Fe, $Fe_3O_4$, $\gamma$-$Fe_2O_3$, or oxy-hydroxides like ferrihydrite, goethite etc) have been detected at least to the detection limit of the X-ray diffractometer used. The peak intensities of the sample annealed at 800°C decreased when compared to the sample annealed at 600°C, indicating crystalline distortion. This could be due to formation of small amount of $MgFe_2O_4$ at such high temperature. The average crystallite size for each annealed samples is calculated from Scherrer formula using FWHM of (214) and (300) peaks after correcting for instrumental broadening and the values are 8±4, 38±5 and 70±6 nm respectively for 400, 600 and 800°C annealed samples.

Based on XRD data, Rietveld refinement has been done on 600 °C annealed sample by MAUD software [13] to further analyze the structural properties and to determine the crystal parameters. XRD diffraction data in 2θ range $30^0$-$110^0$ was used for the Rietveld refinement (Fig. 3). The space group of the *R3c* in the rhombohedral structure where Fe atoms are (x=0, y=0 and z=0.3553) and oxygen atoms are (x=0.3059, y=0 and z=¼) were taken as the staring parameters for the refinement. The crystal parameters of Mg-doped hematite nanoparticles (a=5.0501 Å and c=13.8003 Å) [8] were taken as the initial value and then refined for the intensity matching. The

Rietveld refinement with the experimental data points is shown in figure 3 for the Ca-Mg doped hematite nanoparticles annealed at 600 $^0$C for 4 hours. The lattice parameters are found to be a=b= 5.0494 (5) Å and c= 13.7958(7) Å for the annealed samples as obtained from the Rietveld refinement after intensity matching. These values are slightly lower than the values found for Mg-doped hematite annealed at 400 $^0$C for 2 hours [8].

### 3.3. Mössbauer Spectroscopic Study

$^{57}$Fe Mössbauer spectroscopy is known to be one of the efficient tools to study the local magnetic behavior and the oxidation state of Fe atoms in a particular matrix. The Mössbauer spectra recorded at room temperature for the as-prepared and annealed Ca-Mg co-doped samples have been presented in figure 4(a)-4(d) respectively. The black dots and the solid blue line represent the experimental data points and least square fit of the spectrum respectively. It is observed that the black dots match well with the solid blue line suggesting good fit of each spectrum. The parameters extracted from the fitting of the spectrum are given in Table 2. At the right most column of the table the phases identified from their Mössbauer parameters are given. The Mössbauer parameters from the earlier studies on individually doped hematite are also given in Table 2 for comparison.

The room temperature Mössbauer spectrum of the as prepared sample (figure 4 (a)) can be fitted with a doublet corresponding to isomer shift ≈ 0.32 mm/s and quadrupole splitting ≈ 0.68 mm/s. The parameters are similar to those of ferrihydrite phase [14] though no crystalline phase has been detected from the X-ray diffraction pattern. A doublet can also result from the superparamagnetic phase in these nanosized Ca-Mg co-doped samples if hematite phase is present. The Mössbauer spectrum of as-prepared sample of Mg-doped hematite nanoparticles prepared by the same method had shown magnetic sextet with quadrupole splitting ≈ -0.17 mm/s and hyperfine field ≈ 50.7 T corresponding to crystalline hematite phase along with a doublet of quadrupole splitting ≈ 0.64 mm/s coming from impurity ferrihydrite phase [8]. The Mössbauer spectrum of the sample annealed at 400 $^0$C (shown in figure 4 (b)) shows same kind of behavior as the as-prepared sample with the quadrupole splitting (QS) increasing from 0.68 mm/s to 0.76 mm/s. The presence of superparamagnetism in hematite and/or amorphous ferrihydrite phase can also be observed by very low coercive field in the magnetization measurement discussed below. So, the doublet mentioned in above two samples can come from the superparamagnetic hematite, amorphous ferrihydrite or both phases. In 400 $^0$C annealed sample, ferrihydrite phase is very unlikely to remain which may be present in the as-prepared sample. The Mössbauer spectrum of 400$^o$C annealed Mg-doped hematite sample was fitted with only one sextet with hyperfine field ≈ 51.2 T coming from pure hematite phase [8].

The sample annealed at 600$^o$C is crystallized to pure hematite structure (rhombohedral structure, space group *R3c*) as investigated by Rietveld refinement and has average crystallite size of about 38±5 nm. The spectrum of this sample is shown in figure 4 (c) and can be fitted with a single sextet of isomer shift ≈ 0.35 mm/s, quadrupole splitting ≈ -0.19 mm/s and hyperfine field ≈ 51.2 T corresponding to the pure hematite phase. No indication of any doublet coming from superparamagnetic phase has been found. The spectrum (shown in figure 4 (d)) for further annealed sample (800$^o$C, particle size 70 ± 6 nm) shows small kinks on all the six main peaks. It can be fitted with three magnetic sextets. Sextet with isomer shift ≈ 0.35 mm/s, quadrupole splitting ≈ - 0.19 mm/s and hyperfine field ≈ 51.4 T corresponds to the pure rhombohedral hematite phase. The other two sextets with hyperfine field ≈ 44.6 and 47.9 T corresponds to the spinel $MgFe_2O_4$ phase. The individual fitted spectrum for hematite and $MgFe_2O_4$ phases are shown in red and green lines in figure 4 (d). The negative value of the quadrupole splitting of the pure hematite phase indicates weak ferromagnetic nature of both these samples.

To further analyze the phases in the as-prepared samples we recorded the Mössbauer spectra at 40 K and 20 K. The spectra are presented in figure 5. Both the spectra can be fitted with a paramagnetic doublet (same as room temperature spectrum) with quadrupole splitting (QS) 0.72 mm/s and 0.74 mm/s at 40 K and 20 K respectively. These are slightly higher than the room temperature value 0.68 mm/s. No sextet has been found even at 20K showing that contribution from the superparamagnetic phase, if any, is insignificant and the sample is primarily amorphous ferrihydrite.

### 3.4. Magnetization Measurements

Hematite in bulk form undergoes two types of magnetic transition, the first one at Neel temperature ($T_N$) which is about 950 K [15] below which it behaves like a weak ferromagnetic compound and second one from weak ferromagnetic to antiferromagnetic transition at about 260 K known as Morin transition ($T_M$) [16]. The magnetic properties of hematite particles greatly depend on the particle morphology and sizes of the prepared samples [17]. Ferrihydrite show antiferromagnetic behavior below 120 K with a small ferromagnetic-like moment arising from uncompensated spins present either inside or at the surface of the particles [18].

Room temperature magnetic hysteresis loops of as prepared and three annealed samples are plotted in figure 6(a)-(d) respectively measured upto highest applied field of 1.75 T. The values of magnetization at highest applied magnetic field, remanent magnetization ($M_r$), coercive field ($H_C$) and squareness (S) as calculated from the VSM experiments have been presented in Table 3. The magnetic hysteresis loop for the as-prepared sample looks like typical paramagnetic materials with nearly zero coercive field. The magnetization increases linearly with increasing magnetic field at least upto highest applied magnetic field we have studied (1.75 T). In our previously reported magnetic properties of Mg-doped hematite nanoparticles prepared by same method [8] had shown noticeable hysteresis loop with large coercive field ≈ 1562 Oe due to the presence of spherical nano sized (80nm) hematite particles. The magnetic hysteresis curve for the sample annealed at 400 $^0$C shows some magnetic ordering, as seen from the slope change in the M-H curve, but with nearly zero coercive field. This kind of hysteresis loop is a signature of superparamagnetism. Similar kind of superparamagnetic behavior has been found in 12 and 18 nm sized hematite nanoparticles [19]. For Mg-doped sample annealed at 400 $^0$C hysteresis loop was observed with coercive field ≈ 650 Oe. But here, with co-doping of Ca and Mg the hysteresis is absent with zero coercive field and remanent magnetization. When the particles annealed at 600 $^0$C (particle size 38 nm), the magnetization showed hysteresis and the moments tends to saturate with applied magnetic field. The coercive field is calculated to be ≈ 51 Oe. For particles annealed at 800 $^0$C (particle size ≈ 70 nm), the coercive field increases to ≈ 185 Oe. The magnetization increase as compared with the samples annealed at 600 $^0$C. The increases could be because of small amount of spinel $MgFe_2O_4$ phase present in these nanoparticles as detected from the Mössbauer spectroscopy.

## 4. CONCLUSIONS

In the present study synergistic effect of doping of Ca(II) and Mg(II) on iron oxide/hydroxide phase formation has been reported. With the doping of single ion i.e., Ca(II) or Mg(II), with similar concentrations cubic or spherical hematite particles were formed whereas when both the ions were doped, completely amorphous ferrihydrite was formed. Room temperature, 40K and

20K Mössbauer spectra confirmed that the particles were indeed ferrihydrite as no sextet for $Fe_2O_3$ was observed even at low temperatures. The as prepared sample was annealed at 400, 600 and 800$^o$C. The phase formation at various temperatures, were confirmed from XRD and Mössbauer studies. The coercivity values increase from nearly zero to 185.25 Oe with the increase in annealing temperature from 400 to 800$^o$C.

**References**

**1.** J.-M. Tulliani, C. Baroni, C. Lopez and L. Dessemond, J. Eur. Ceram. Soc., 31, 2357 (2011).

**2.** J.-M. Tulliani and P. Bonville, Ceram. Int. 31, 507 (2005).

**3.** A. Bak, W. Choi and H. Park, Appl. Catal. B: Environmental, 110, 207 (2011).

**4.** X. Lian, X. Yang, S. Liu, Y. Xu, C. Jiang, J. Chen and R. Wang, Appl. Surf. Sci. 258, 2307 (2011).

**5.** M. D. S. Ramos, M. D. S. Santos, L. P. Gomes, A. Albornoz and M. D. C. Rangel, Appl. Catal. A: General, 341, 12 (2008).

**6.** M. Mohapatra, Chandan Upadhyay, Brejesh, S.Anand, R .P. Das and H. C. Verma, J. Mag. Mag. Mat. 295, 44 (2005).

**7.** M. Mohapatra, D. Behera, S. Layek, S. Anand, H .C. Verma and B.K. Mishra, Cryst. Growth Des. 12, 18 (2012).

**8.** M. Mohapatra, Samar Layek, S. Anand, H. C. Verma and B. K. Mishra (submitted).

**9.** A. I. Vogel, A Text Book of Quantitative Inorganic Analysis, English Language Book Society and Longmans Green Publishers, London (2000).

**10.** M. Mohapatra, T. Padhi, T. Dash, P. Singh, S. Anand and B. K. Mishra, Toxicological and Environmental Chemistry, 93, 844 (2011).

11. M. Mohapatra, K. Rout, P. Singh, S. Anand, S. Layek, H. C. Verma and B .K. Mishra, J. Hazard. Mat. 186, 1751 (2011).

12. K. Baltpurvins, R. Burns and G. Lawrance, Environ. Sci. Technol. 31, 1024 (1997).

13. L. Lutterotti, MAUD, Version 2.046, 2006, (http://www.ing.unitn.it/maud/).

14. T. S. Berquo, J. J. Erbs, A. Lindquist, R. L. Penn and S. K. Banerjee, J. Phys.: Condens. Mat. 21, 176005 (2009).

15. C. G. Shull, W. A. Strauser and E. O. Wollan, Phys. Rev. 83, 333(1951).

16. F.J. Morin, Phys. Rev. 78, 819 (1950).

17. S. Mitra, S. Das, K. Mandal and S. Chaudhuri, Nanotechnology 18, 275608 (2007).

18. R. S. Zergenyi, A. M. Hirt, S. Zimmermann, J.P. Dobson and W. Lowrie, J. Geophys. Res. 105, 8297 (2000).

19. J. Jacob and M. A. Khadar, J. Mag. Mag. Mat. 322, 614 (2010).

**Table 1.** Effect of surfactant mediation in presence/absence of Ca(II) and/or (Mg(II) while maintaining rest of conditions same.

| pptn. conditions | Chemical analysis | | | phases | Ref |
| --- | --- | --- | --- | --- | --- |
| | % Fe | % Mg | %Ca | | |
| No CTAB, No Mg 100 mL 1M Fe(III), NaOH neutralizing agent pH 10 | 55.83% | - | - | 6-line ferrihydrite | [10] |
| 5mL 10%CTAB, No Mg 100 mL 1M Fe(III), NaOH neutralizing agent, pH 10 | 60.13 | - | - | mixed ferrihydrite, α FeOOH, α-Fe$_2$O$_3$ | [11] |
| 5 mL 10%CTAB, 0.05M Mg(II), 100 mL 1M Fe(III), NaOH neutralizing agent, pH 10 | 63.8 | 1.53 | - | α-Fe$_2$O$_3$ major, ferrihydrite minor(11%) | [8] |
| 5 mL 10%CTAB, 0.05M Ca(II), 100 mL 1M Fe(III), NaOH neutralizing agent, pH 10 | 66.8 | - | 2.25 | α-Fe$_2$O$_3$ major, ferrihydrite minor(8%) | [7] |
| 5 mL 10%CTAB, 0.05M Ca(II) + 0.05M Mg(II)), 100 mL 1M Fe(III), NaOH neutralizing agent, pH 10 | 55.83 | 1.17 | 1.87 | amorphous ferrihydrite* | [present] |

*XRD results discussed below

**Table 2** Mössbauer parameters i.e., isomer shift (IS), quadrupole splitting (QS), hyperfine field ($B_{HF}$), line width (LWD) and area ratio of the phases indicated at the right most columns of as-prepared and annealed Ca-Mg co-doped samples.

| Sample | Annealing | T (K) | IS (mm/s) | QS (mm/s) | LWD (mm/s) | $B_{HF}$ (T) | Area (%) | Phase | Reference |
|---|---|---|---|---|---|---|---|---|---|
| Ca(II) doped | as- prepared | 300 K | 0.22 | 0.66 | 0.41 | - | 8 | Ferrihydrite | [7] |
| | | | 0.35 | -0.17 | 0.32 | 50.8 | 92 | Hematite | |
| Mg(II) doped | as- prepared | 300 K | 0.26 | 0.64 | 0.43 | - | 11 | Ferrihydrite | [8] |
| | | | 0.35 | -0.17 | 0.33 | 50.7 | 89 | Hematite | |
| Ca-Mg doped | as- prepared | 300 K | 0.32 | 0.68 | 0.48 | - | 100 | Ferrihydrite | Present Study |
| | | 40 K | 0.40 | 0.72 | 0.62 | - | 100 | Ferrihydrite | |
| | | 20 K | 0.40 | 0.74 | 0.61 | - | 100 | Ferrihydrite | |
| | 400 °C | 300 K | 0.31 | 0.76 | 0.54 | - | 100 | Hematite | |
| | 600 °C | 300 K | 0.35 | -0.19 | 0.31 | 51.2 | 100 | Hematite | |
| | 800 °C | 300 K | 0.35 | -0.19 | 0.28 | 51.4 | 82 | Hematite | |
| | | | 0.27 | 0.03 | 0.42 | 44.6 | 10 | $MgFe_2O_4$ | |
| | | | 0.37 | -0.02 | 0.33 | 47.9 | 8 | | |

**Table 3** Magnetization at highest applied magnetic field (1.75T) ($M_S$), remanent magnetization ($M_r$), coercive field ($H_C$) and squareness (S) of as prepared and annealed samples measured at 300K using vibrating sample magnetometer (VSM).

| Sample | Annealing | $M_S$ (emu/g) | $M_r$ (emu/g) | $H_C$ (Oe) | S= $M_r/M_S$ | Reference |
|---|---|---|---|---|---|---|
| Ca(II) doped | as-prepared | 0.9287 | 0.1159 | 1254.9 | 0.125 | [7] |
| Mg(II) doped | as-prepared | 1.061 | 0.133 | 1561.66 | 0.126 | [8] |
| Ca-Mg doped | as-prepared | 2.729 | ≈ 0 | ≈ 0 | ≈ 0 | Present Study |
| | 400 $^0$C | 2.576 | ≈ 0 | ≈ 0 | ≈ 0 | |
| | 600 $^0$C | 2.432 | 0.0641 | 51.24 | 0.026 | |
| | 800 $^0$C | 3.039 | 0.6755 | 185.25 | 0.222 | |

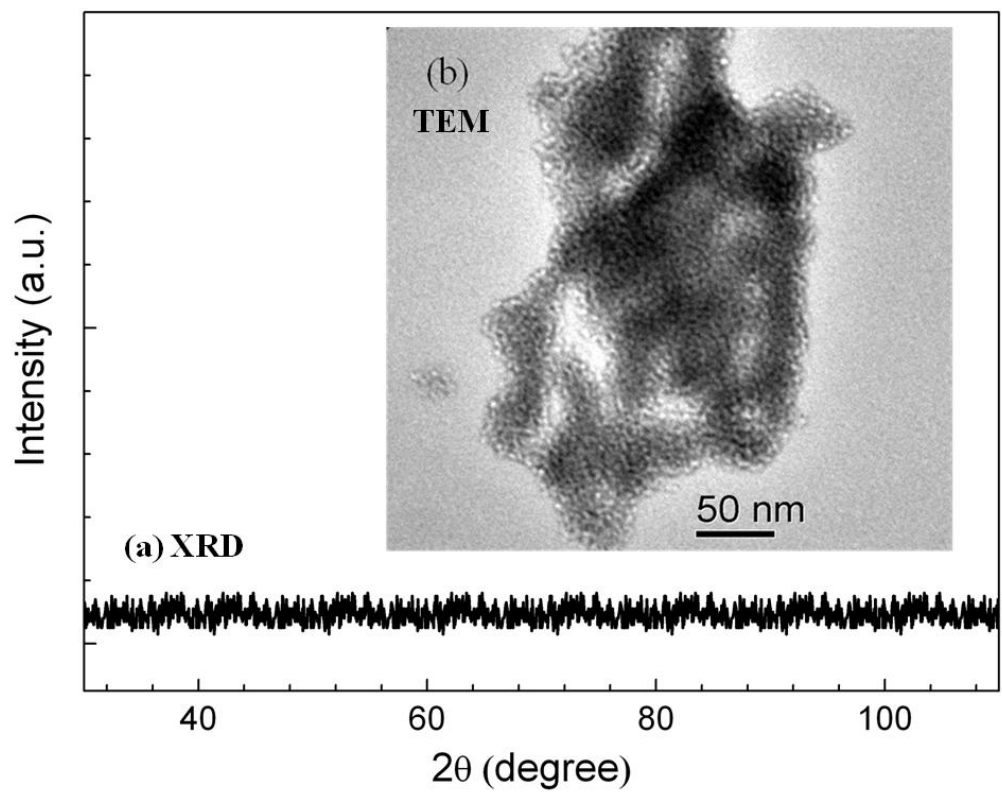

**Figure 1** (a) X-ray diffraction (XRD) pattern and (b) Transmission Electron Microscopy (TEM) image of as prepared Ca-Mg doped sample.

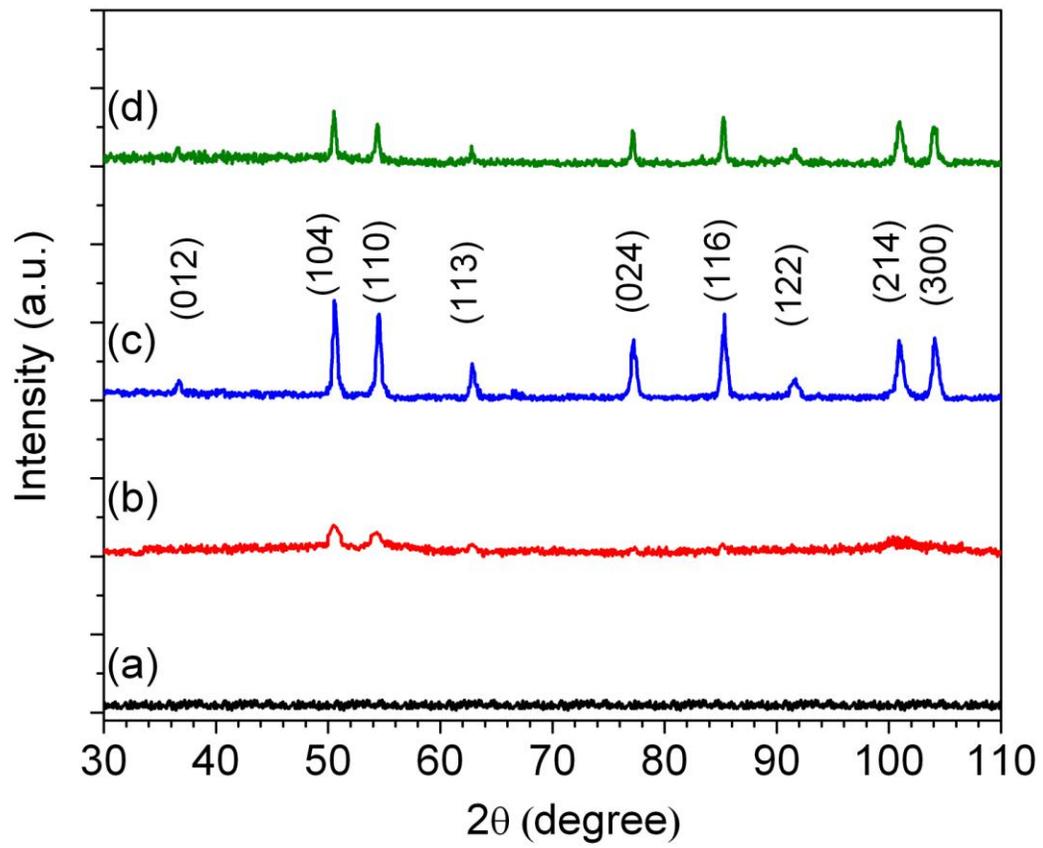

**Figure 2** Room temperature X-ray diffraction patterns of (a) as-prepared, (b) 400, (c) 600 and (d) 800 $^0$C annealed samples.

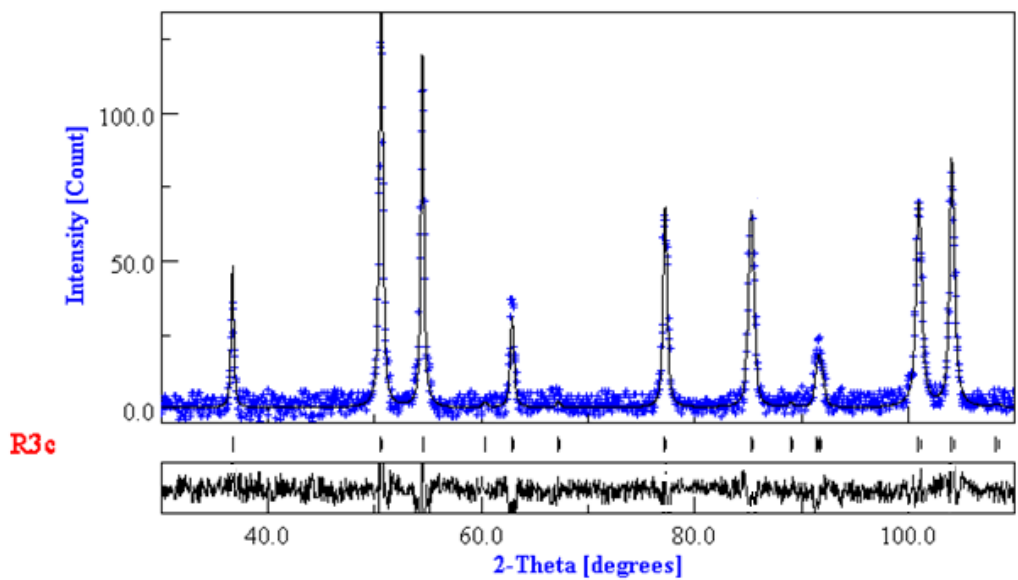

**Figure 3** Rietveld refinement profile of the room temperature powder XRD pattern of Ca-Mg co-doped hematite sample annealed at 600 $^0$C using MAUD software [13]. Blue dots are the experimental points, solid black is fit, in-between black bars are Bragg position of the *R3c* phase and difference between experimental data and theoretical fit are shown at the bottom.

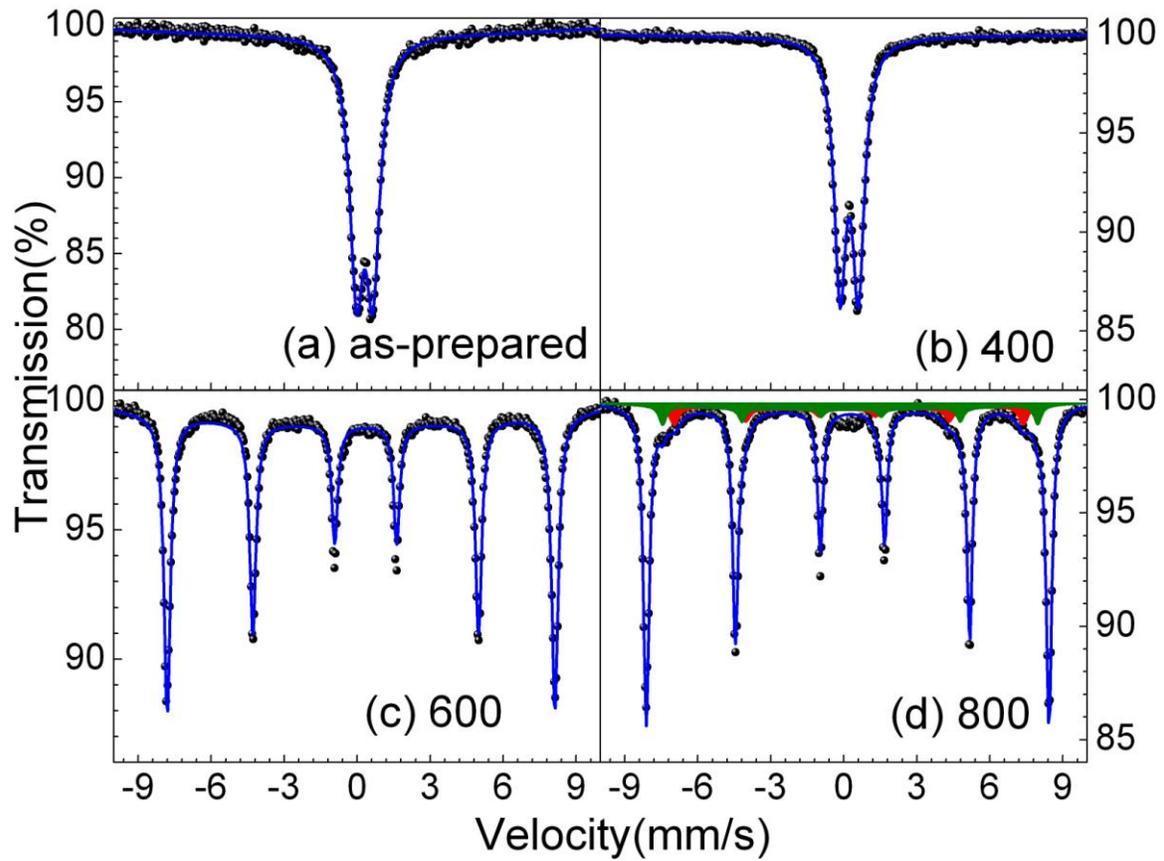

**Figure 4** Mössbauer spectrum for (a) as-prepared, (b) 400, (c) 600 and (d) 800 $^0$C annealed Ca-Mg co-doped samples measured at 300 K. Black spheres are the experimental data points and solid blue line is the Lorentzian fit.

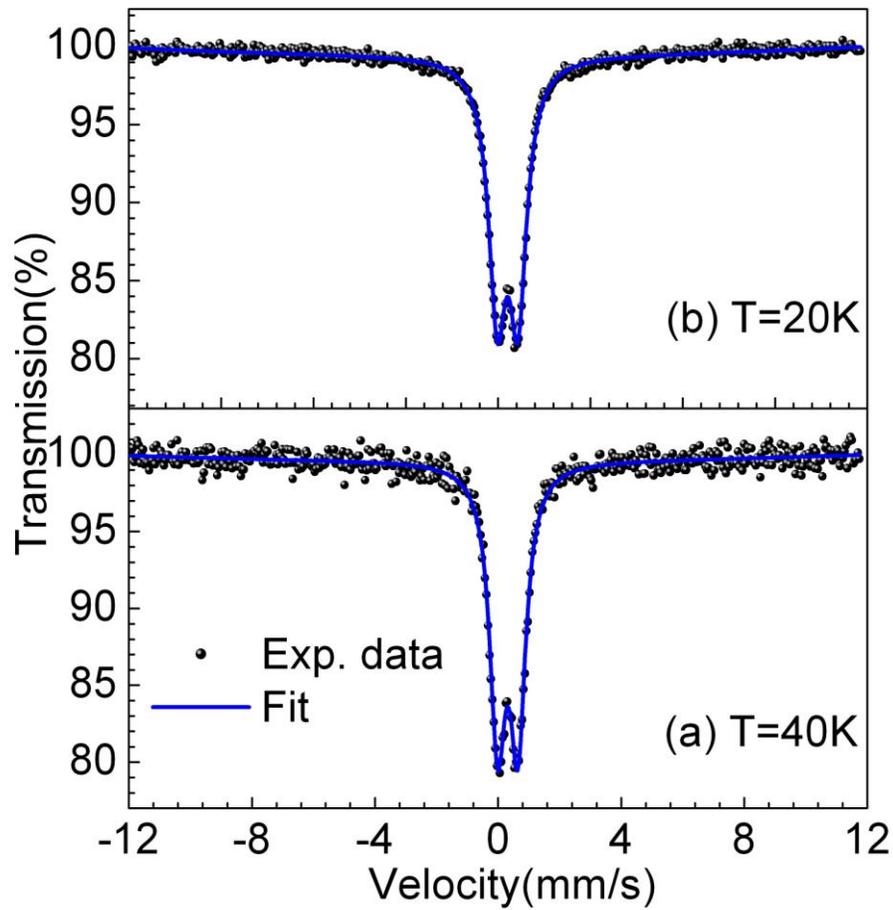

**Figure 5** Mössbauer spectrum of as prepared samples at (a) 40 K and (b) 20 K.

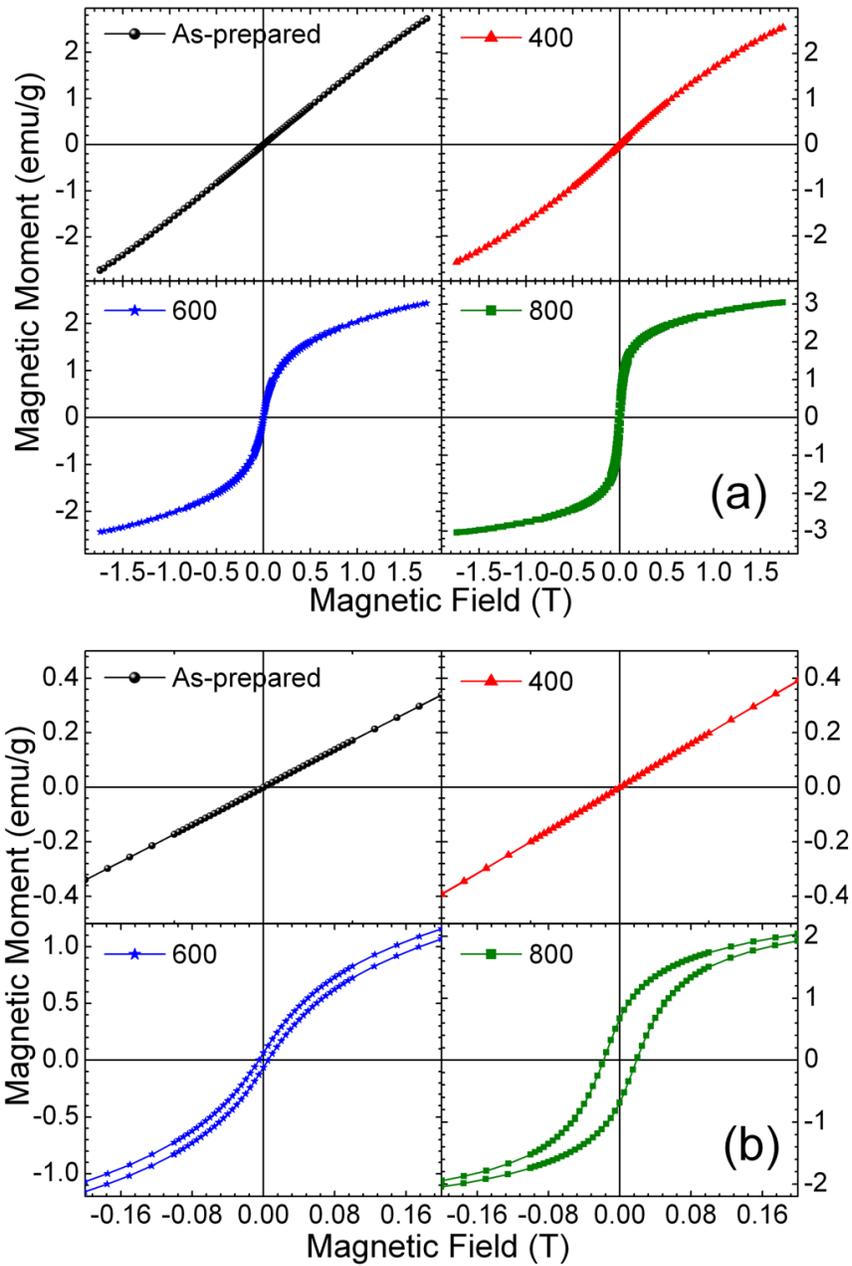

**Fig. 6.** (a) Magnetization as a function of applied field (hysteresis curves) of Ca-Mg co-doped as-prepared ferrihydrite and annealed (different temperature) samples measured using vibrating sample magnetometer (VSM) at 300 K and (b) same loops in low field region.